\documentclass[aps,prd,twocolumn,groupedaddress,amssymb,eqsecnum,showpacs,epsfig]{revtex4}
\usepackage{graphicx}
\usepackage{bm}
\usepackage{dcolumn}
\usepackage{amsmath}
\def \lleq {\lower0.9ex\hbox{ $\buildrel < \over \sim$} ~}
\def \ggeq {\lower0.9ex\hbox{ $\buildrel > \over \sim$} ~}

\def \omms   {\Omega_m}

\def \omm  {\Omega_{0 {\rm m}}}

\def \beq  {\begin{equation}}
\def \eeq  {\end{equation}}
\def \ber  {\begin{eqnarray}}
\def \eer  {\end{eqnarray}}

\begin{document}
\newcommand{\newc}{\newcommand}

\newc{\be}{\begin{equation}}
\newc{\ee}{\end{equation}}
\newc{\ba}{\begin{eqnarray}}
\newc{\ea}{\end{eqnarray}}
\newc{\bea}{\begin{eqnarray*}}
\newc{\eea}{\end{eqnarray*}}
\newc{\D}{\partial}
\newc{\ie}{{\it i.e.} }
\newc{\eg}{{\it e.g.} }
\newc{\etc}{{\it etc.} }
\newc{\etal}{{\it et al.}}
\newc{\lcdm }{$\Lambda$CDM }
\newcommand{\nn}{\nonumber}
\newc{\ra}{\rightarrow}
\newc{\lra}{\leftrightarrow}
\newc{\lsim}{\buildrel{<}\over{\sim}}
\newc{\gsim}{\buildrel{>}\over{\sim}}
\title{Testing $\Lambda$CDM with the Growth Function $\delta(a)$: Current Constraints}
\author{S. Nesseris and L. Perivolaropoulos}
\affiliation{Department of Physics, University of Ioannina, Greece}
\date{\today}

\begin{abstract}
We have compiled a dataset consisting of 22 datapoints at a redshift
range (0.15,3.8) which can be used to constrain the linear
perturbation growth rate $f=\frac{d\ln\delta}{d\ln a}$. Five of
these data-points constrain directly the growth rate $f$ through
either redshift distortions or change of the power spectrum with
redshift. The rest of the datapoints constrain $f$ indirectly
through the rms mass fluctuation $\sigma_8(z)$ inferred from
Ly-$\alpha$ at various redshifts. Our analysis tests the consistency
of the $\Lambda$CDM model and leads to a constraint of the
Wang-Steinhardt growth index $\gamma$ (defined from
$f=\Omega_m(a)^\gamma$) as $\gamma=0.67^{+0.20}_{-0.17}$. This
result is clearly consistent at $1\sigma$ with the value
$\gamma=\frac{6}{11}=0.545$ predicted by $\Lambda$CDM. We also apply
our analysis on a new null test of $\Lambda$CDM which is similar to
the one recently proposed by Chiba and Nakamura (arXiv:0708.3877)
but does not involve derivatives of the expansion rate $H(z)$. This
also leads to the fact that $\Lambda$CDM provides an excellent fit
to the current linear growth data.
\end{abstract}
%
%
\maketitle

\section{Introduction}
Most of the current observational evidence for the accelerating
expansion of the universe and the existence of dark energy comes
from geometrical tests that measure directly the integral of the
expansion rate of the universe $H(z)$ at various redshifts. Such
tests include measurements of the luminosity distance by using
standard candles like type Ia supernovae (SnIa) \cite{SN} and
measurement of the angular luminosity distance using standard rulers
(last scattering horizon scale \cite{CMB}, baryon acoustic
oscillations peak \cite{BAO} etc). Even though these tests are
presently the most accurate probe of dark energy, the mere
determination of the expansion rate $H(z)$ is not able to provide
significant insight into the properties of dark energy and
distinguish it from models that attribute the accelerating expansion
to modifications of general relativity. The additional observational
input that is required is the growth function $\delta(z)\equiv
\frac{\delta\rho}{\rho}(z)$ of the linear matter density contrast as
a function of redshift. The combination of the observed functions
$H(z)$ and $\delta(z)$ can provide significant insight into the
properties of dark energy (e.g. sound speed, existence of
anisotropic stress etc) or even distinguish it from modified gravity
theories\cite{Boisseau:2000pr}.

The observational description of the expansion rate $H(z)$ is
usually made through the use of parameterizations of the effective
equation of state of dark energy $w(z)\equiv
\frac{p_{de}(z)}{\rho_{de}(z)}$ which is related to $H(z)$ as \be
w(z)=\frac{\frac{2}{3}(1+z)\frac{dlnH}{dz} -
1}{1-\left(\frac{H_0}{H} \right)^2 \omm (1+z)^3} \label{whz} \ee The
most commonly used such parametrization is the CPL parametrization
\cite{Chevallier:2000qy,Linder:2002et} \be w(z)=w_0+w_1\frac{z}{1+z}
\label{cpl} \ee The cosmological constant (\lcdm ) corresponds to
parameter values $w_0=-1$, $w_1=0$ and is not only consistent with
all current geometrical tests but it is also favored
\cite{CMB,Sullivan:2007pd} by most of them compared to other
parameter values.

The corresponding parametrization of the linear growth function
$\delta(z)$ can be made efficiently by introducing a growth index
$\gamma$ defined by \be \frac{d\ln\delta(a)}{d\ln
a}=\omms(a)^\gamma \label{gamdef} \ee where $a=\frac{1}{1+z}$ is
the scale factor and \be \omms(a)\equiv \frac{H_0^2 \omm
a^{-3}}{H^2(a)} \label{ommsdef} \ee This parametrization was
originally introduced by Wang and Steinhardt \cite{Wang:1998gt}
(see also \cite{Linder:2007hg} for more recent discussions) and
was shown to provide an excellent fit to
$\frac{d\ln\delta(a)}{d\ln a}$ corresponding to various general
relativistic cosmological models for specific values of $\gamma$.
In particular, it was shown that for dark energy with slowly
varying $w(z)\simeq w_0$ the parameter $\gamma$ in a flat universe
is \be \gamma=\frac{3(w_0-1)}{6w_0-5} \label{gwrel} \ee which for
the \lcdm  case ($w=-1$) reduces to $\gamma=\frac{6}{11}$. It is
therefore clear that the observational determination of the growth
index $\gamma$ can be used to test \lcdm.

The observational determination of the growth index $\gamma$
requires knowledge not only of $\delta(z)$ but also of $H(z)$ and
$\omm$ (see equation (\ref{ommsdef})). It is possible however to
construct more direct tests of \lcdm  which require knowledge only
of $\delta(z)$ and $H(z)$. Such a null test of \lcdm  was recently
proposed in Ref. \cite{Chiba:2007vm} where it was suggested that the
validity of \lcdm  requires the following equality of observables
\be \frac{(H(z)^2/H_0^2)'}{(1+z)^2 \delta'(0)^2}\int_0^\infty
\frac{\delta(z)\delta'(z)}{(1+z)} dz +1=0 \label{cntest} \ee where
$'\equiv \frac{d}{dz}$. In fact, as discussed section III, there is
an improved version of this null test that does not involve
derivative of the observable $H(z)$ and therefore it is less prone
to observational errors.

Both of the \lcdm  tests discussed above (the growth index
$\gamma$ and the null test) require observational determination of
$\delta(z)$. There are several observational approaches that can
lead to the determination of $\delta(z)$. For example, redshift
distortions of galaxy power spectra \cite{Hawkins:2002sg}, the rms
mass fluctuation $\sigma_8(z)$ inferred from galaxy and
$Ly-\alpha$ surveys at various redshifts
\cite{Viel:2004bf}-\cite{Viel:2005ha}, weak lensing statistics
\cite{Kaiser:1996tp}, baryon acoustic oscillations
\cite{Seo:2003pu}, X-ray luminous galaxy clusters
\cite{Mantz:2007qh}, Integrated Sachs-Wolfe (ISW) effect
\cite{Pogosian:2005ez} etc. Unfortunately, the currently available
data are limited in number and accuracy and come mainly from the
first two categories. They involve significant error bars and
non-trivial assumptions that hinder a reliable determination of
$\delta(z)$. In addition, a large part of the available data are
at high redshifts ($z>1$) where \lcdm is degenerate with most
other dark energy models since dark energy is subdominant compared
to matter at high redshifts in most models.

Nevertheless, it is still instructive to consider the presently
available data to investigate the possible weak constraints that
can be imposed on \lcdm. Such a task serves two purposes
\begin{enumerate} \item It can be used as a {\it paradigm} for the time
when more accurate data will be available \item It can provide
constraints from a {\it dynamical} test which are orthogonal and
completely independent from the usual geometrical tests.
\end{enumerate} Thus, in what follows we use a wide range of
presently available data on both redshift distortions and
$\sigma_8(z)$ to determine the observed growth index $\gamma$ and
test \lcdm  in two ways \begin{enumerate} \item Comparing the
measured value of $\gamma$ with the \lcdm  prediction
$\gamma=\frac{6}{11}$ \item Implementing a new null test that
exploits the consistency between $H(z)$ and $\delta(z)$ in the
context of \lcdm. \end{enumerate} The structure of this paper is
the following: In the next section we present the dataset we have
compiled and use it to determine the best fit value of $\gamma$
under the assumption of a \lcdm  background. In section III we
derive a new null test for \lcdm  and apply it using the fit
performed in section II. Finally, in section IV we conclude,
summarize and present the future prospects of the present study.

\section{Fitting the Growth Index}

According to general relativity, the equations that determine the
evolution of the density contrast $\delta$ in a flat background
consisting of matter with density $\rho_m$ and dark energy with
$\rho_{de}=\frac{p_{de}}{w}$ are of the form \ba &&{\ddot \delta} +
2
\frac{\dot a}{a} {\dot \delta} = 4\pi G \rho_m \delta \label{del-t}\\
&&\left(\frac{\dot a}{a}\right)^2=\frac{8\pi G}{3}\left(\rho_m +
\rho_{de}\right) \label{fe1} \\
&&2\frac{\ddot a}{a}+\left(\frac{\dot a}{a}\right)^2=-8\pi G w
\rho_{de} \label{fe2}\ea It is straightforward to change variables
in eq. (\ref{del-t}) from $t$ to $\ln a$
($\frac{d}{dt}=H\frac{d}{d\ln a}$) to obtain \cite{Wang:1998gt} \be
(\ln \delta)''+(\ln \delta)'^2 + (\ln
\delta)'\left[\frac{1}{2}-\frac{3}{2}w(1-\omms
(a))\right]=\frac{3}{2} \omms (a) \label{lnda} \ee where we used
(\ref{fe2}) and \be \omms (a)\equiv \frac{\rho_m(a)}{\rho_m (a) +
\rho_{de} (a)} \label{omadef} \ee as in (\ref{ommsdef}). A further
change of variables from $\ln a$ to $\omms (a)$ can be made by
considering the differential of eq. (\ref{omadef}) and using energy
conservation ($d\rho=-3(\rho+p)d\ln a$) leads to \be d\omms=3w\omms
(1-\omms)d\ln a \label{domda} \ee Using (\ref{domda}) in
(\ref{lnda}) we get \cite{Wang:1998gt} \be 3w\omms
(1-\omms)\frac{df}{d\omms} + f^2 +f\left[\frac{1}{2}-\frac{3}{2} w
(1-\omms)\right]=\frac{3}{2}\omms \label{fomdiff} \ee where we have
set \be f\equiv \frac{d\ln \delta}{d\ln a} \label{fdef} \ee Using
the ansatz \be f=\omms^{\gamma(\omms)} \label{fomgam} \ee in eq.
(\ref{fomdiff}) and expanding around $\omms=1$ (good approximation
especially at $z\gsim 1$) we find to lowest order \be
\gamma=\frac{3(w-1)}{6w-5} \label{gam-w} \ee which reduces to
$\gamma=\frac{6}{11}$ for \lcdm ($w=-1$).

Equations (\ref{fomgam}) and (\ref{gam-w}) provide excellent
approximations to the numerically obtained form of $f(z)$. This is
demonstrated in Fig. 1 where we plot the numerically obtained
solution of eq. (\ref{lnda}) for the normalized growth \be
g(z)\equiv \frac{\delta (z)}{\delta(0)} \label{gzdef} \ee in the
case of \lcdm ($\omm=0.3$) along with the corresponding
approximate result \be g(z) = e^{\int_1^{\frac{1}{1+z}}\omms
(a)^\gamma \frac{da}{a} } \label{gzom} \ee with
$\gamma=\frac{6}{11}$ obtained from \be
f(\omm,\gamma,a)=a\frac{d\delta/da}{\delta} = \omms (a)^\gamma
\label{fomgam1} \ee The difference between the two approaches is
less than $0.1 \% $.

Despite of the impressive agreement between numerical result and
analytical approximation, there has been a recent attempt
\cite{Polarski:2007rr} to improve further on the analytical
approximation by considering an expansion of $\gamma$ in redshift
space up to first order in $z$ \be \gamma =  \gamma_0 +\gamma_0' z
\label{gamexp}\ee where $\gamma_0$ and $\gamma_0'$ are constants.
Using (\ref{gamexp}) in (\ref{fomgam}) and (\ref{fomdiff}) it may
be shown that \cite{Polarski:2007rr} \ber \gamma_0'&=&
(\frac{1}{\ln\omms}) \\ \nn
&\cdot&\left[\omms^{\gamma_0}+3(\gamma_0-\frac{1}{2})w_0(1-\omms)
- \frac{3}{2}\omms^{1-\gamma_0}+\frac{1}{2}\right] \label{gamprim}
\eer which for $\gamma_0=\frac{6}{11}$, $w_0=-1$, $\omms=0.3$
becomes $\gamma_0'=-0.012$. Since $\gamma_0'$ is very small for
\lcdm we set it to zero in what follows and assume
$\gamma=\gamma_0 = constant$. In the Appendix however, we
generalize our fit using equation (\ref{gamexp}) and show that the
introduction of the new parameter $\gamma_0'$ increases
significantly the errors and the allowed parameter region at
$1\sigma$.

\begin{figure}[!t]
\hspace{0pt}\rotatebox{0}{\resizebox{.5\textwidth}{!}{\includegraphics{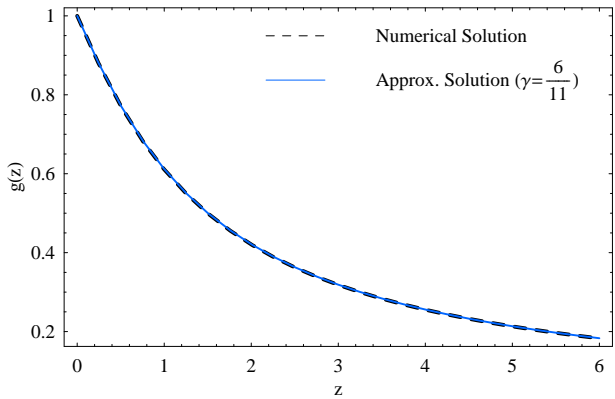}}}
\vspace{0pt}{\caption{The numerically obtained solution of eq.
(\ref{lnda}) for the normalized growth of eq. (\ref{gzdef}) in the
case of \lcdm ($\omm=0.3$) (black dashed line) along with the
corresponding approximate result with $\gamma=\frac{6}{11}$
obtained from eq. (\ref{fomgam1}) (blue continuous line). The
agreement between the two approaches is excellent.}} \label{fig1}
\end{figure}

Our goal in this section is to fit the parameter $\gamma$ using
observational data and compare it with the value
$\gamma=\frac{6}{11}$ of \lcdm. The most useful currently
available data that can be used to constrain $\delta(z)$ (and
$\gamma$) involve the redshift distortion parameter $\beta$
\cite{Hamilton:1997zq} observed through the anisotropic pattern of
galactic redshifts on cluster scales (for a pedagogical discussion
see \cite{Nesseris:2006er}). The parameter $\beta$ is related to
the growth rate $f$ as \be \beta = \frac{d\ln \delta/d\ln
a}{b}=\frac{f}{b} \label{betf} \ee where $b$ is the bias factor
connecting total matter perturbations $\delta$ and galaxy
perturbations $\delta_g$ ($b=\frac{\delta_g}{\delta}$).

The currently available data for the parameters $\beta$ and $b$ at
various redshifts are shown in Table I along with the inferred
growth rates and references. This is an expanded version of the
dataset used in Ref. \cite{Di Porto:2007ym} where a similar analysis
was performed using a different parametrization suitable for
modified gravity models. Notice that the Refs. of Table I have
assumed \lcdm (with $\omm=0.3$) when converting redshifts to
distances for the power spectra and therefore their use to test
models different from \lcdm may not be reliable. In addition, as
pointed out in Ref. \cite{Wang:2007ht}, the points of Table I
obtained from Refs \cite{Tegmark:2006az}, \cite{Ross:2006me},
\cite{daAngela:2006mf} correspond to measurements of $\beta(z)$ but
the estimate of the bias $b$ is made by comparing numerically
simulated power spectra with the observed ones. Since the
numerically simulated power spectra have assumed a \lcdm cosmology,
the resulting $f_{obs}=\beta\; b$ should be interpreted carefully
can only be used to test the consistency of \lcdm cosmology. Given
this ambiguity we have evaluated the best fit value of the index
$\gamma$ both with and without the three points discussed above (see
equations (\ref{gambf}) and (\ref{gambf2})).

\vspace{0pt}
\begin{table}[!t]
\begin{center}
\caption{The currently available data for the parameters $\beta$
and $b$ at various redshifts along with the inferred growth rates
and references. Notice that Ref. \cite{McDonald:2004xn} only
reports the growth rate and not the $\beta$ and $b$ parameters
since the growth rate was obtained directly from the change of
power spectrum $Ly-\alpha$ forest data in SDSS at various redshift
slices. \label{table1}}
\begin{tabular}{ccccc}
\hline \hline\\
\hspace{6pt}   $z$  &\hspace{6pt} $\beta$ & \hspace{6pt} $b$ &\hspace{6pt}  $f_{obs}$ &\hspace{6pt}  \textbf{Ref.} \\
\hline\\
\hspace{6pt}   0.15 &\hspace{6pt} $0.49 \pm 0.09$         &\hspace{6pt} $1.04\pm 0.11$ &\hspace{6pt}  $0.51 \pm 0.11$ & \cite{Hawkins:2002sg},\cite{Verde:2001sf}  \\
\hspace{6pt}   0.35 &\hspace{6pt} $0.31 \pm 0.04$         &\hspace{6pt} $2.25\pm 0.08$ &\hspace{6pt}  $0.70 \pm 0.18$ & \cite{Tegmark:2006az}  \\
\hspace{6pt}   0.55 &\hspace{6pt} $0.45 \pm 0.05$         &\hspace{6pt} $1.66\pm 0.35$ &\hspace{6pt}  $0.75 \pm 0.18$ & \cite{Ross:2006me}  \\
\hspace{6pt}   1.4  &\hspace{6pt} $0.60 ^{+0.14}_{-0.11}$ &\hspace{6pt} $1.5 \pm 0.20$ &\hspace{6pt}  $0.90 \pm 0.24$ & \cite{daAngela:2006mf}  \\
\hspace{6pt}   3.0  &\hspace{6pt} $-$                     &\hspace{6pt}$-$             &\hspace{6pt}  $1.46 \pm 0.29$ & \cite{McDonald:2004xn} \\
\hline \hline \\
\end{tabular}
\end{center}
\end{table}

In the same Table we also show the growth rate obtained in Ref.
\cite{McDonald:2004xn} which does not rely on $\beta$ but is based
on a different strategy, namely by finding directly the change of
power spectrum $Ly-\alpha$ forest data in SDSS at various redshift
slices. This point (in addition to the first one) has been used
previously by other authors (see Ref. \cite{Di Porto:2007ym}) in a
similar context.

Using the data of Table I we can perform a maximum likelihood
analysis in order to find $\gamma$ and check its consistency with
the \lcdm  value $\frac{6}{11}$. We thus construct \be \chi_f^2
(\omm,\gamma) = \sum_i
\left[\frac{f_{obs}(z_i)-f_{th}(z_i,\gamma)}{\sigma_{f_{obs}}}\right]^2
\label{chif2} \ee where $f_{obs}$ and $\sigma_{fobs}$ are obtained
from Table I while $f_{th}(z_i,\gamma)$ is obtained from eq.
(\ref{fomgam}).

An alternative observational probe of the growth function
$\delta(z)$ is the redshift dependence of the rms mass fluctuation
$\sigma_8 (z)$ defined by \be \sigma^2(R,z)=\int_0^\infty
W^2(kR)\Delta^2(k,z)\frac{dk}{k} \label{sigrz} \ee with \ba
W(kR)&=&3\left(\frac{\sin (kR)}{(kR)^3}-\frac{\cos
(kR)}{(kR)^2}\right) r \label{wkr} \\ \Delta^2 (kz)&=&4\pi k^3
P_\delta (k,z) \label{delkz} \ea with $R=8h^{-1}Mpc$ and $P_\delta
(k,z)$ the mass power spectrum at redshift $z$. The function
$\sigma_8 (z)$ is connected with $\delta (z)$ as \be
\sigma_8(z)=\frac{\delta(z)}{\delta(0)} \sigma_8(z=0)
\label{s8del} \ee which implies \be s_{th}(z_1,z_2)\equiv
\frac{\sigma_8(z_1)}{\sigma_8(z_2)}=\frac{\delta(z_1)}{\delta(z_2)}=\frac{e^{\int_1^{\frac{1}{1+z_1}}\omms^\gamma
(a)\frac{da}{a}}} {e^{\int_1^{\frac{1}{1+z_2}}\omms^\gamma
(a)\frac{da}{a}}}\label{sthrats8} \ee where we made use of eq.
(\ref{gzom}). Most of the currently available datapoints $\sigma_8
(z_i)$ originate from the observed redshift evolution of the flux
power spectrum of $Ly-\alpha$ forest
\cite{Viel:2004bf,Viel:2005ha,vimos}. These datapoints are shown
in Table II along with the corresponding reference sources. Notice
that the data from Ref. \cite{vimos} were obtained using the
normalized by $a$ growth factor $\delta$ and in our analysis we
taken this into account. These data are not as useful as the
redshift distortion factors for the determination of $\gamma$ for
two reasons
\begin{enumerate} \item The rms fluctuation $\sigma_8 (z)$ is not
connected directly with the growth rate $f(z)$. Instead, it is
related with $f(z)$ through the integral of eq. (\ref{gzom}).
\item Most of the $Ly-\alpha$ $\sigma_8$ data appear at high
redshifts where \lcdm  is degenerate with most other dark energy
models.\end{enumerate}

\vspace{0pt}
\begin{table}[!t]
\begin{center}
\caption{The currently available data for the rms fluctuation
$\sigma_8 (z)$ at various redshifts and references. Notice that
the data from Ref. \cite{vimos} were obtained using the normalized
by $a$ growth factor $\delta$ and in our analysis we took this
into account. \label{table2}}
\begin{tabular}{cccc}
\hline \hline\\
\hspace{6pt}   $z$  &\hspace{6pt}  $\sigma_8 $ & $\sigma_{\sigma_8 }$ &\hspace{6pt}  \textbf{Ref.} \\
\hline\\
\hspace{6pt}   2.125 &\hspace{6pt}  0.95 &\hspace{6pt}  0.17 & \cite{Viel:2004bf}  \\
\hspace{6pt}   2.72 &\hspace{6pt}  0.92 &\hspace{6pt}  0.17 &   \\
\hline\\
\hspace{6pt}   2.2  &\hspace{6pt}  0.92 &\hspace{6pt}  0.16 & \cite{Viel:2005ha}  \\
\hspace{6pt}   2.4  &\hspace{6pt}  0.89 &\hspace{6pt}  0.11 &   \\
\hspace{6pt}   2.6  &\hspace{6pt}  0.98 &\hspace{6pt}  0.13 &   \\
\hspace{6pt}   2.8  &\hspace{6pt}  1.02 &\hspace{6pt}  0.09 &   \\
\hspace{6pt}   3.0  &\hspace{6pt}  0.94 &\hspace{6pt}  0.08 &   \\
\hspace{6pt}   3.2  &\hspace{6pt}  0.88 &\hspace{6pt}  0.09 &   \\
\hspace{6pt}   3.4  &\hspace{6pt}  0.87 &\hspace{6pt}  0.12 &   \\
\hspace{6pt}   3.6  &\hspace{6pt}  0.95 &\hspace{6pt}  0.16 &   \\
\hspace{6pt}   3.8  &\hspace{6pt}  0.90 &\hspace{6pt}  0.17 &   \\
\hline\\
\hspace{6pt}   0.35 &\hspace{6pt}  0.55 &\hspace{6pt}  0.10 & \cite{vimos}  \\
\hspace{6pt}   0.6  &\hspace{6pt}  0.62 &\hspace{6pt}  0.12 &   \\
\hspace{6pt}   0.8  &\hspace{6pt}  0.71 &\hspace{6pt}  0.11 &   \\
\hspace{6pt}   1.0  &\hspace{6pt}  0.69 &\hspace{6pt}  0.14 &   \\
\hspace{6pt}   1.2  &\hspace{6pt}  0.75 &\hspace{6pt}  0.14 &   \\
\hspace{6pt}   1.65 &\hspace{6pt}  0.92 &\hspace{6pt}  0.20 &   \\
\hline \hline \\
\end{tabular}
\end{center}
\end{table}

Using the data of Table II we construct the corresponding $\chi_s^2$
defined as \be \chi_s^2 (\omm,\gamma)=\sum_i
\left[\frac{s_{obs}(z_i,z_{i+1})-s_{th}(z_i,z_{i+1})}{\sigma_{s_{obs,i}}}\right]^2
\label{chi2s} \ee where $\sigma_{s_{obs,i}}$ is derived by error
propagation from the corresponding $1\sigma$ errors of $\sigma_8
(z_i)$ and $\sigma_8 (z_{i+1})$ while $s_{th}(z_i,z_{i+1})$ is
defined in eq. (\ref{sthrats8}). We can thus construct the combined
$\chi_{tot}^2 (\omm,\gamma)$ as \be \chi_{tot}^2(\omm,\gamma)\equiv
\chi_{f}^2(\omm,\gamma)+\chi_{s}^2(\omm,\gamma) \label{chitot2} \ee
Notice however that all Refs report their data points having assumed
a \lcdm model when converting redshifts to distances with $\omm=0.3$
(except \cite{daAngela:2006mf} that used $\omm=0.25$ and
\cite{Viel:2004bf,Viel:2005ha} that used $\omm=0.26$).

Setting $\omm=0.3$ and minimizing $\chi_{tot}^2$ with respect to
$\gamma$ we find \be \gamma=0.674^{+0.195}_{-0.169}
\label{gambf}\ee which differs somewhat from the corresponding
result of Ref. \cite{Di Porto:2007ym} because we have used a
broader dataset, a different parametrization for $f$ and we have
assumed \lcdm  as our fiducial model thus avoiding the
marginalization of the parameter $w_0$. The result (\ref{gambf})
indicates that the \lcdm value of $\gamma=\frac{6}{11}=0.545$ is
well within $1\sigma$ from the best fit and is clearly consistent
with data. The imposed constraints however are rather weak and
even a flat model with matter only (SCDM) predicting $\gamma=0.6$
(set $w=0$ in eq. (\ref{gam-w})) is consistent with the data. 
Also, this result indicates that if it was only the value of
$f_{obs}$ and the measured $\beta$ that would be required then it
could have been obtained trivially from the $\Lambda$CDM $f_{obs}$
(which is analytically known) and the measured $\beta$ without need
for a simulation. In that case we would also have found almost
perfect agreement with $\Lambda$CDM. Instead we find a somewhat
larger value of $\gamma$. Performing the same analysis but by
excluding the three datapoints at $z=0.35,0.55$ and $1.4$ yields a
slightly different value for \be \gamma=0.73 \pm 0.23 \label{gambf2}
\ee but again well within $1\sigma$ from the $\Lambda$CDM value.

Ignoring the $\sigma_8 (z)$ data of Table II leads to a negligible
change in the best fit of all 22 points to $\gamma=0.663\pm 0.2$.
This is consistent with the above discussion on the usefulness of
these data. Alternatively, assuming the \lcdm value for $\gamma$
($\gamma=\frac{6}{11}$) and minimizing with respect to $\omm$ we
find $\omm=0.24^{+0.09}_{-0.07}$.

The cosmological data for the growth rate $f(z)$ are shown in Fig.2
along with the best theoretical fit $f=\omms(z)^\gamma$ with
$\omm=0.3$ and the corresponding $1\sigma$ errors (shaded region).
In the same plot we show (dashed line) the corresponding $f_{\Lambda
CDM}(z)$ obtained by solving numerically eq. (\ref{lnda}) for $w=-1$
and $\omm=0.3$.  Clearly, the best fit (continuous line)  shows a
minor difference from \lcdm (dashed line) only at low redshifts. We
therefore conclude that \lcdm is consistent with current data for
the growth factor $\delta(z)$ and the consistency is maximized for
$\omm=0.24^{+0.09}_{-0.07}$.

\begin{figure}[!t]
\hspace{0pt}\rotatebox{0}{\resizebox{.5\textwidth}{!}{\includegraphics{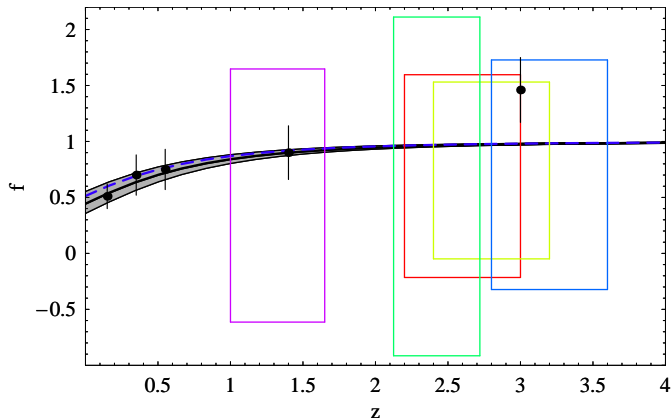}}}
\vspace{0pt}{\caption{The cosmological data for the growth rate
$f(z)$ along with the best theoretical fit $f=\omms(z)^\gamma$
with $\omm=0.3$ (black continuous line) and the corresponding
$1\sigma$ errors (shaded region). The errorboxes on f are obtained
using the ratios at the specific redshifts. Clearly, the best fit
shows a minor difference from \lcdm (blue dashed line) only at low
redshifts. }} \label{fig2}
\end{figure}

\section{Alternative Tests of \lcdm  using $\delta(z)$}

The test of \lcdm  discussed in the previous section requires prior
knowledge of the parameter $\omm$ since the growth function data are
not accurate enough to produce a simultaneous fit for both $\omm$
and $\gamma$ with reasonably small errors. Thus it is useful to
derive a test that depends only on the observables $H(z)$ and
$\delta(z)$. Such a consistency test has been recently discussed in
Ref. \cite{Chiba:2007vm} and involves both $H(z)$, $\delta(z)$ and
their derivatives (see eq. (\ref{cntest})). Here we derive an
improved version of this test that is independent of the derivative
of $H(z)$ and therefore it is less subject to observational errors.
We start from eq. (\ref{del-t}) and change variables from $t$ to $a$
to obtain \be
\frac{dH(a)^2}{da}+2\left(\frac{3}{a}+\frac{\delta''}{\delta'}\right)H^2=\frac{3\omm
H_0^2 \delta}{a^5 \delta'} \label{dadiff} \ee where $'\equiv
\frac{d}{da}$. The solution to (\ref{dadiff}) is
\cite{Starobinsky:1998fr} \be
\frac{H^2}{H_0^2}=\frac{3\omm}{a^6\delta'(a)^2}\int_0^{a}a\delta(a)\delta'(a)da
\label{hdelsola} \ee which may be expressed in redshift space as \be
\frac{H^2}{H_0^2}=-\frac{3\omm
(1+z)^2}{\delta'(z)^2}\int_z^{\infty}\frac{\delta(z)\delta'(z)}{1+z}
dz \label{hdelsolz} \ee and setting $z=0$ we find \be \omm =
-\frac{1}{3}\delta'(0)^2 \left[\int_0^\infty
\frac{\delta(z)\delta'(z)}{1+z} dz\right]^{-1} \label{ommdyn} \ee In
order to promote eq. (\ref{ommdyn}) to a null test for \lcdm we must
express $\omm$ in terms of geometrical observables like $H(z)$.

\begin{figure}[!t]
\hspace{0pt}\rotatebox{0}{\resizebox{.5\textwidth}{!}{\includegraphics{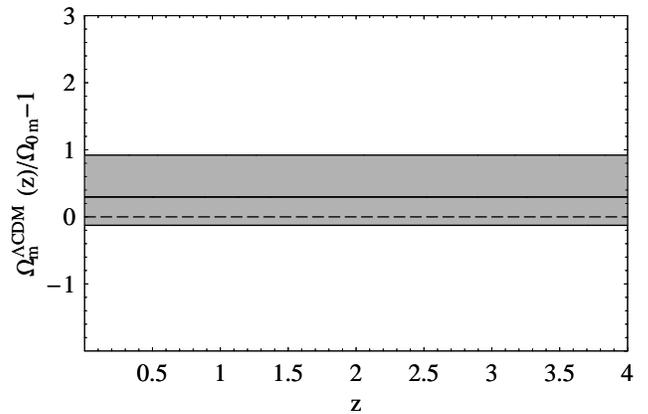}}}
\vspace{0pt}{\caption{ The $1\sigma$ range (shaded region) of the
lhs of eq. (\ref{nulltest}) (black continuous line). Notice that by
construction it is independent of the redshift $z$ since we have
assumed that the geometric part of eq. (\ref{nulltest}) ($H(z)$)
behaves like \lcdm ($\omm=0.3$). The value $0$ corresponding to
\lcdm for both $H(z)$ and $\delta(z)$ (dashed line) is clearly well
within $1\sigma$ from the best fit (continuous line). This is to be
expected since the range of $\gamma$ in eq. (\ref{gambf}) includes
the value $\gamma=\frac{6}{11}$. }} \label{fig3}
\end{figure}

In the context of \lcdm  we have \be \omms^{\Lambda CDM}
(z)=\left[\left(\frac{H(z)}{H_0}\right)^2-1\right]\frac{1}{(1+z)^3-1}
\label{om0lam} \ee leading to $\omms^{\Lambda CDM} (z)=\omm$ when
$H(z)$ has the \lcdm form. By dividing (\ref{om0lam}) with
(\ref{ommdyn}) we have \be \frac{\omms^{\Lambda
CDM}(z)}{\omm}-1=-\frac{3 \left(\frac{H(z)^2}{H_0^2}-1\right)
\int_0^{\infty}\frac{\delta(z)\delta'(z)}{1+z}
dz}{\left[(1+z)^3-1\right]\delta'(0)^2}-1=0 \label{nulltest} \ee
where the last equality should hold only if \lcdm  is a valid
theory. Using (\ref{gzom}), (\ref{fomgam1}) and (\ref{gambf}) in
(\ref{nulltest}) we can find the $1\sigma$ range of the lhs of eq.
(\ref{nulltest}) which is shown in Fig. 3 (shaded region). Notice
that by construction it is independent of the redshift $z$ since we
have assumed that the geometric part of eq. (\ref{nulltest})
($H(z)$) behaves like \lcdm  ($\omm=0.3$). The value $0$
corresponding to \lcdm  is clearly well within $1\sigma$ as shown in
Fig. 3. This is to be expected since the range of $\gamma$ in eq.
(\ref{gambf}) includes the value $\gamma=\frac{6}{11}$. Had we
allowed for a more general form of $H(z)$ then the $1\sigma$ range
of eq. (\ref{nulltest}) would be redshift dependent. However, as
discussed in section II, in that case the allowed range of $\gamma$
would be less reliable since the surveys leading to the data of
Tables I and II convert redshifts to distances using \lcdm with
$\omm=0.3$. Therefore, these datapoints can only be used to test
\lcdm.

\begin{figure*}[!t]
\hspace{0pt}\rotatebox{0}{\resizebox{1\textwidth}{!}{\includegraphics{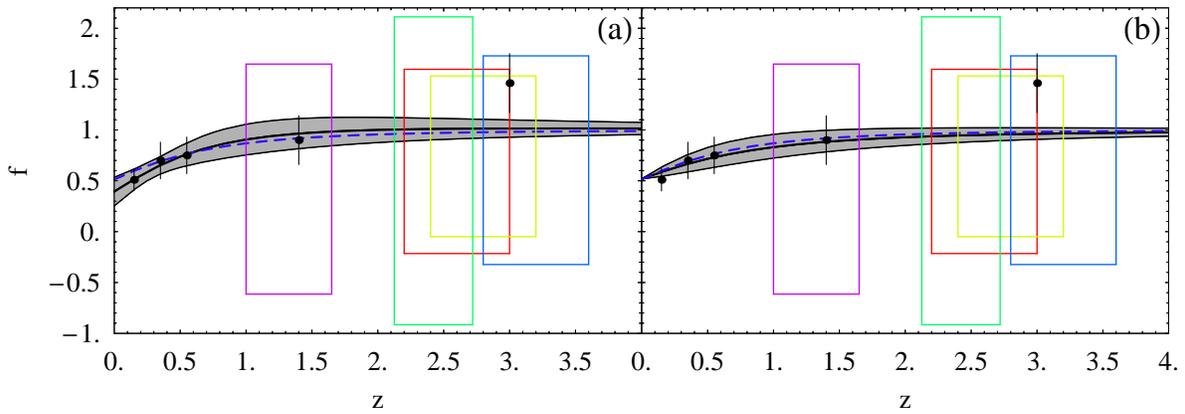}}}
\vspace{0pt}{\caption{ Same as Fig. 2 with a generalized
parametrization (\ref{gamexp}) where either both parameters
$\gamma_0$, $\gamma_0'$ are allowed to vary (a) or only
$\gamma_0'$ is allowed to vary while $\gamma_0$ is fixed to its
\lcdm value (b).}} \label{fig3}
\end{figure*}

\section{Conclusion}
We have compiled a dataset consisting of redshift distortion
factors $\beta (z)$ and rms mass fluctuations $\sigma_8 (z)$ at
various redshifts obtained from galaxy and $Ly-\alpha$ forest
redshift surveys. Using this dataset we have obtained the best fit
form of the linear growth function $\delta(z)$ using the Wang and
Steinhardt parametrization (\ref{fomgam1}) with $\omm=0.3$. We
have found a best fit value of $\gamma$ as
$\gamma=0.674^{+0.195}_{-0.169}$, a range which includes the \lcdm
value $\gamma=\frac{6}{11}$. Thus \lcdm is in excellent agreement
with current linear growth data. We have reached the same
conclusion by applying a generalized version of the null test of
Ref. \cite{Chiba:2007vm}.

The combination of geometrical ($H(z)$) and dynamical ($\delta(z)$)
observables used here to test \lcdm could also be used to test
modified gravity theories which can not be easily tested by using
geometrical tests alone. However, in that case, care should be taken
to reanalyze the power spectra using the proper form of $H(z)$ when
converting redshifts to distances.

Nevertheless, given the current uncertainties in the growth rate
observations, data of much better quality will be needed in order
to distinguish between \lcdm and modified gravity theories. These
data will most likely come from large scale weak lensing surveys
like DUNE, which is expected to measure the the equation of state
of dark energy to a precision better than $5\%$
\cite{Refregier:2006vt}.

{\bf Numerical Analysis:} The mathematica files with the numerical
analysis of this study may be found at
http://nesseris.physics.uoi.gr/growth/growth.htm or may be sent by
e-mail upon request.

{\bf Acknowledgements:} We thank Y. Wang for useful comments. This
work was supported by the European Research and Training Network
MRTPN-CT-2006 035863-1 (UniverseNet). S.N. acknowledges support from
the Greek State Scholarships Foundation (I.K.Y.).

\section{Appendix}
It is straightforward to generalize our analysis of section II
using (\ref{gamexp}) in (\ref{fomgam1}), (\ref{chif2}),
(\ref{sthrats8}). In that case we have two parameters to fit and
we obtain \ba
\gamma_0 &=&0.77 \pm 0.29 \\
\gamma_0' &=&-0.38 \pm 0.85 \ea Clearly, the $1\sigma$ error region
is significantly increased in that case (see Fig. 4) due to the
introduction of the additional parameter and \lcdm
($\gamma_0=\frac{6}{11}$, $\gamma_0' \simeq 0$) remains well within
$1\sigma$ from the best fit.

In order to avoid the increased error region we may also set
$\gamma_0=\frac{6}{11}$ and fit $\gamma_0'$ only. In this case we
get \be \gamma_0'=0.17\pm 0.54 \ee which is again consistent with
\lcdm with more reasonable errors.

\end{document}